\catcode`@=11
\newif\ifams
\amstrue
%
%
%
%
\newfam\bdifam
\newfam\bsyfam
\newfam\bssfam
\newfam\msafam
\newfam\msbfam
\newif\ifxxpt   
\newif\ifxviipt 
\newif\ifxivpt  
\newif\ifxiipt  
\newif\ifxipt   
\newif\ifxpt    
\newif\ifixpt   
\newif\ifviiipt 
\newif\ifviipt  
\newif\ifvipt   
\newif\ifvpt    
%
%
\def\headsize#1#2{\def\headb@seline{#2}%
                \ifnum#1=20\def\he@d{twenty}%
                           \def\smhe@d{twelve}%
                           \def\vshe@d{nine}%
                           \ifxxpt\else\xdef\f@ntsize{\he@d}%
                           \def\m@g{4}\def\s@ze{20.74}%
                           \loadheadfonts\xxpttrue\fi
                           \ifxiipt\else\xdef\f@ntsize{\smhe@d}%
                           \def\m@g{1}\def\s@ze{12}%
                           \loadxiiptfonts\xiipttrue\fi
                           \ifixpt\else\xdef\f@ntsize{\vshe@d}%
                           \def\s@ze{9}%
                           \loadsmallfonts\ixpttrue\fi
                      \else
                \ifnum#1=17\def\he@d{seventeen}%
                           \def\smhe@d{eleven}%
                           \def\vshe@d{eight}%
                           \ifxviipt\else\xdef\f@ntsize{\he@d}%
                           \def\m@g{3}\def\s@ze{17.28}%
                           \loadheadfonts\xviipttrue\fi
                           \ifxipt\else\xdef\f@ntsize{\smhe@d}%
                           \loadxiptfonts\xipttrue\fi
                           \ifviiipt\else\xdef\f@ntsize{\vshe@d}%
                           \def\s@ze{8}%
                           \loadsmallfonts\viiipttrue\fi
                      \else\def\he@d{fourteen}%
                           \def\smhe@d{ten}%
                           \def\vshe@d{seven}%
                           \ifxivpt\else\xdef\f@ntsize{\he@d}%
                           \def\m@g{2}\def\s@ze{14.4}%
                           \loadheadfonts\xivpttrue\fi
                           \ifxpt\else\xdef\f@ntsize{\smhe@d}%
                           \def\s@ze{10}%
                           \loadxptfonts\xpttrue\fi
                           \ifviipt\else\xdef\f@ntsize{\vshe@d}%
                           \def\s@ze{7}%
                           \loadviiptfonts\viipttrue\fi
                \ifnum#1=14\else\message{Header size should be 20, 17 or 14 point
                              will now default to 14pt}\fi
                \fi\fi\headfonts}
%
%
\def\textsize#1#2{\def\textb@seline{#2}%
                 \ifnum#1=12\def\t@xt{twelve}%
                           \def\smt@xt{eight}%
                           \def\vst@xt{six}%
                           \ifxiipt\else\xdef\f@ntsize{\t@xt}%
                           \def\m@g{1}\def\s@ze{12}%
                           \loadxiiptfonts\xiipttrue\fi
                           \ifviiipt\else\xdef\f@ntsize{\smt@xt}%
                           \def\s@ze{8}%
                           \loadsmallfonts\viiipttrue\fi
                           \ifvipt\else\xdef\f@ntsize{\vst@xt}%
                           \def\s@ze{6}%
                           \loadviptfonts\vipttrue\fi
                      \else
                \ifnum#1=11\def\t@xt{eleven}%
                           \def\smt@xt{seven}%
                           \def\vst@xt{five}%
                           \ifxipt\else\xdef\f@ntsize{\t@xt}%
                           \def\s@ze{11}%
                           \loadxiptfonts\xipttrue\fi
                           \ifviipt\else\xdef\f@ntsize{\smt@xt}%
                           \loadviiptfonts\viipttrue\fi
                           \ifvpt\else\xdef\f@ntsize{\vst@xt}%
                           \def\s@ze{5}%
                           \loadvptfonts\vpttrue\fi
                      \else\def\t@xt{ten}%
                           \def\smt@xt{seven}%
                           \def\vst@xt{five}%
                           \ifxpt\else\xdef\f@ntsize{\t@xt}%
                           \loadxptfonts\xpttrue\fi
                           \ifviipt\else\xdef\f@ntsize{\smt@xt}%
                           \def\s@ze{7}%
                           \loadviiptfonts\viipttrue\fi
                           \ifvpt\else\xdef\f@ntsize{\vst@xt}%
                           \def\s@ze{5}%
                           \loadvptfonts\vpttrue\fi
                \ifnum#1=10\else\message{Text size should be 12, 11 or 10 point
                              will now default to 10pt}\fi
                \fi\fi\textfonts}
%
%
\def\smallsize#1#2{\def\smallb@seline{#2}%
                 \ifnum#1=10\def\sm@ll{ten}%
                           \def\smsm@ll{seven}%
                           \def\vssm@ll{five}%
                           \ifxpt\else\xdef\f@ntsize{\sm@ll}%
                           \loadxptfonts\xpttrue\fi
                           \ifviipt\else\xdef\f@ntsize{\smsm@ll}%
                           \def\s@ze{7}%
                           \loadviiptfonts\viipttrue\fi
                           \ifvpt\else\xdef\f@ntsize{\vssm@ll}%
                           \def\s@ze{5}%
                           \loadvptfonts\vpttrue\fi
                       \else
                 \ifnum#1=9\def\sm@ll{nine}%
                           \def\smsm@ll{six}%
                           \def\vssm@ll{five}%
                           \ifixpt\else\xdef\f@ntsize{\sm@ll}%
                           \def\s@ze{9}%
                           \loadsmallfonts\ixpttrue\fi
                           \ifvipt\else\xdef\f@ntsize{\smsm@ll}%
                           \def\s@ze{6}%
                           \loadviptfonts\vipttrue\fi
                           \ifvpt\else\xdef\f@ntsize{\vssm@ll}%
                           \def\s@ze{5}%
                           \loadvptfonts\vpttrue\fi
                       \else
                           \def\sm@ll{eight}%
                           \def\smsm@ll{six}%
                           \def\vssm@ll{five}%
                           \ifviiipt\else\xdef\f@ntsize{\sm@ll}%
                           \def\s@ze{8}%
                           \loadsmallfonts\viiipttrue\fi
                           \ifvipt\else\xdef\f@ntsize{\smsm@ll}%
                           \def\s@ze{6}%
                           \loadviptfonts\vipttrue\fi
                           \ifvpt\else\xdef\f@ntsize{\vssm@ll}%
                           \def\s@ze{5}%
                           \loadvptfonts\vpttrue\fi
                 \ifnum#1=8\else\message{Small size should be 10, 9 or 
                            8 point will now default to 8pt}\fi
                \fi\fi\smallfonts}
\def\F@nt{\expandafter\font\csname}
\def\Sk@w{\expandafter\skewchar\csname}
\def\@nd{\endcsname}
\def\@step#1{ scaled \magstep#1}
\def\@half{ scaled \magstephalf}
\def\@t#1{ at #1pt}
%
%
\def\loadheadfonts{\bigf@nts
\F@nt \f@ntsize bdi\@nd=cmmib10 \@t{\s@ze}%
\Sk@w \f@ntsize bdi\@nd='177
\F@nt \f@ntsize bsy\@nd=cmbsy10 \@t{\s@ze}%
\Sk@w \f@ntsize bsy\@nd='60
\F@nt \f@ntsize bss\@nd=cmssbx10 \@t{\s@ze}}
%
%
\def\loadxiiptfonts{\bigf@nts
\F@nt \f@ntsize bdi\@nd=cmmib10 \@step{\m@g}%
\Sk@w \f@ntsize bdi\@nd='177
\F@nt \f@ntsize bsy\@nd=cmbsy10 \@step{\m@g}%
\Sk@w \f@ntsize bsy\@nd='60
\F@nt \f@ntsize bss\@nd=cmssbx10 \@step{\m@g}}
%
%
\def\loadxiptfonts{%
\font\elevenrm=cmr10 \@half
\font\eleveni=cmmi10 \@half
\skewchar\eleveni='177
\font\elevensy=cmsy10 \@half
\skewchar\elevensy='60
\font\elevenex=cmex10 \@half
\font\elevenit=cmti10 \@half
\font\elevensl=cmsl10 \@half
\font\elevenbf=cmbx10 \@half
\font\eleventt=cmtt10 \@half
\ifams\font\elevenmsa=msam10 \@half
\font\elevenmsb=msbm10 \@half\else\fi
\font\elevenbdi=cmmib10 \@half
\skewchar\elevenbdi='177
\font\elevenbsy=cmbsy10 \@half
\skewchar\elevenbsy='60
\font\elevenbss=cmssbx10 \@half}
%
%
\def\loadxptfonts{%
\font\tenbdi=cmmib10
\skewchar\tenbdi='177
\font\tenbsy=cmbsy10 
\skewchar\tenbsy='60
\ifams\font\tenmsa=msam10 
\font\tenmsb=msbm10\else\fi
\font\tenbss=cmssbx10}%
%
%
\def\loadsmallfonts{\smallf@nts
\ifams
\F@nt \f@ntsize ex\@nd=cmex\s@ze
\else
\F@nt \f@ntsize ex\@nd=cmex10\fi
\F@nt \f@ntsize it\@nd=cmti\s@ze
\F@nt \f@ntsize sl\@nd=cmsl\s@ze
\F@nt \f@ntsize tt\@nd=cmtt\s@ze}
%
%
\def\loadviiptfonts{%
\font\sevenit=cmti7
\font\sevensl=cmsl8 at 7pt
\ifams\font\sevenmsa=msam7 
\font\sevenmsb=msbm7
\font\sevenex=cmex7
\font\sevenbsy=cmbsy7
\font\sevenbdi=cmmib7\else
\font\sevenex=cmex10
\font\sevenbsy=cmbsy10 at 7pt
\font\sevenbdi=cmmib10 at 7pt\fi
\skewchar\sevenbsy='60
\skewchar\sevenbdi='177
\font\sevenbss=cmssbx10 at 7pt}%
%
%
\def\loadviptfonts{\smallf@nts
\ifams\font\sixex=cmex7 at 6pt\else
\font\sixex=cmex10\fi
\font\sixit=cmti7 at 6pt}
%
%
\def\loadvptfonts{%
\font\fiveit=cmti7 at 5pt
\ifams\font\fiveex=cmex7 at 5pt
\font\fivebdi=cmmib5
\font\fivebsy=cmbsy5
\font\fivemsa=msam5 
\font\fivemsb=msbm5\else
\font\fiveex=cmex10
\font\fivebdi=cmmib10 at 5pt
\font\fivebsy=cmbsy10 at 5pt\fi
\skewchar\fivebdi='177
\skewchar\fivebsy='60
\font\fivebss=cmssbx10 at 5pt}
\def\bigf@nts{%
\F@nt \f@ntsize rm\@nd=cmr10 \@step{\m@g}%
\F@nt \f@ntsize i\@nd=cmmi10 \@step{\m@g}%
\Sk@w \f@ntsize i\@nd='177
\F@nt \f@ntsize sy\@nd=cmsy10 \@step{\m@g}%
\Sk@w \f@ntsize sy\@nd='60
\F@nt \f@ntsize ex\@nd=cmex10 \@step{\m@g}%
\F@nt \f@ntsize it\@nd=cmti10 \@step{\m@g}%
\F@nt \f@ntsize sl\@nd=cmsl10 \@step{\m@g}%
\F@nt \f@ntsize bf\@nd=cmbx10 \@step{\m@g}%
\F@nt \f@ntsize tt\@nd=cmtt10 \@step{\m@g}%
\ifams
\F@nt \f@ntsize msa\@nd=msam10 \@step{\m@g}%
\F@nt \f@ntsize msb\@nd=msbm10 \@step{\m@g}\else\fi}
\def\smallf@nts{%
\F@nt \f@ntsize rm\@nd=cmr\s@ze
\F@nt \f@ntsize i\@nd=cmmi\s@ze 
\Sk@w \f@ntsize i\@nd='177
\F@nt \f@ntsize sy\@nd=cmsy\s@ze
\Sk@w \f@ntsize sy\@nd='60
\F@nt \f@ntsize bf\@nd=cmbx\s@ze 
\ifams
\F@nt \f@ntsize bdi\@nd=cmmib\s@ze 
\F@nt \f@ntsize bsy\@nd=cmbsy\s@ze 
\F@nt \f@ntsize msa\@nd=msam\s@ze 
\F@nt \f@ntsize msb\@nd=msbm\s@ze
\else
\F@nt \f@ntsize bdi\@nd=cmmib10 \@t{\s@ze} 
\F@nt \f@ntsize bsy\@nd=cmbsy10 \@t{\s@ze}\fi 
\Sk@w \f@ntsize bdi\@nd='177
\Sk@w \f@ntsize bsy\@nd='60
\F@nt \f@ntsize bss\@nd=cmssbx10 \@t{\s@ze}}%
%
%
\def\headfonts{%
\textfont0=\csname\he@d rm\@nd        
\scriptfont0=\csname\smhe@d rm\@nd
\scriptscriptfont0=\csname\vshe@d rm\@nd
\def\rm{\fam0\csname\he@d rm\@nd
\def\sc{\csname\smhe@d rm\@nd}}%
\textfont1=\csname\he@d i\@nd         
\scriptfont1=\csname\smhe@d i\@nd
\scriptscriptfont1=\csname\vshe@d i\@nd
\textfont2=\csname\he@d sy\@nd        
\scriptfont2=\csname\smhe@d sy\@nd
\scriptscriptfont2=\csname\vshe@d sy\@nd
\textfont3=\csname\he@d ex\@nd        
\scriptfont3=\csname\smhe@d ex\@nd
\scriptscriptfont3=\csname\smhe@d ex\@nd
\textfont\itfam=\csname\he@d it\@nd   
\scriptfont\itfam=\csname\smhe@d it\@nd
\scriptscriptfont\itfam=\csname\vshe@d it\@nd
\def\it{\fam\itfam\csname\he@d it\@nd
\def\sc{\csname\smhe@d it\@nd}}%
\textfont\slfam=\csname\he@d sl\@nd   
\def\sl{\fam\slfam\csname\he@d sl\@nd
\def\sc{\csname\smhe@d sl\@nd}}%
\textfont\bffam=\csname\he@d bf\@nd   
\scriptfont\bffam=\csname\smhe@d bf\@nd
\scriptscriptfont\bffam=\csname\vshe@d bf\@nd
\def\bf{\fam\bffam\csname\he@d bf\@nd
\def\sc{\csname\smhe@d bf\@nd}}%
\textfont\ttfam=\csname\he@d tt\@nd   
\def\tt{\fam\ttfam\csname\he@d tt\@nd}%
\textfont\bdifam=\csname\he@d bdi\@nd 
\scriptfont\bdifam=\csname\smhe@d bdi\@nd
\scriptscriptfont\bdifam=\csname\vshe@d bdi\@nd
\def\bdi{\fam\bdifam\csname\he@d bdi\@nd}%
\textfont\bsyfam=\csname\he@d bsy\@nd 
\scriptfont\bsyfam=\csname\smhe@d bsy\@nd
\def\bsy{\fam\bsyfam\csname\he@d bsy\@nd}%
\textfont\bssfam=\csname\he@d bss\@nd 
\scriptfont\bssfam=\csname\smhe@d bss\@nd
\scriptscriptfont\bssfam=\csname\vshe@d bss\@nd
\def\bss{\fam\bssfam\csname\he@d bss\@nd}%
\ifams
\textfont\msafam=\csname\he@d msa\@nd 
\scriptfont\msafam=\csname\smhe@d msa\@nd
\scriptscriptfont\msafam=\csname\vshe@d msa\@nd
\textfont\msbfam=\csname\he@d msb\@nd 
\scriptfont\msbfam=\csname\smhe@d msb\@nd
\scriptscriptfont\msbfam=\csname\vshe@d msb\@nd
\else\fi
\normalbaselineskip=\headb@seline pt%
\setbox\strutbox=\hbox{\vrule height.7\normalbaselineskip 
depth.3\baselineskip width0pt}%
\def\sc{\csname\smhe@d rm\@nd}\normalbaselines\rm}
%
%
\def\textfonts{%
\textfont0=\csname\t@xt rm\@nd        
\scriptfont0=\csname\smt@xt rm\@nd
\scriptscriptfont0=\csname\vst@xt rm\@nd
\def\rm{\fam0\csname\t@xt rm\@nd
\def\sc{\csname\smt@xt rm\@nd}}%
\textfont1=\csname\t@xt i\@nd         
\scriptfont1=\csname\smt@xt i\@nd
\scriptscriptfont1=\csname\vst@xt i\@nd
\textfont2=\csname\t@xt sy\@nd        
\scriptfont2=\csname\smt@xt sy\@nd
\scriptscriptfont2=\csname\vst@xt sy\@nd
\textfont3=\csname\t@xt ex\@nd        
\scriptfont3=\csname\smt@xt ex\@nd
\scriptscriptfont3=\csname\smt@xt ex\@nd
\textfont\itfam=\csname\t@xt it\@nd   
\scriptfont\itfam=\csname\smt@xt it\@nd
\scriptscriptfont\itfam=\csname\vst@xt it\@nd
\def\it{\fam\itfam\csname\t@xt it\@nd
\def\sc{\csname\smt@xt it\@nd}}%
\textfont\slfam=\csname\t@xt sl\@nd   
\def\sl{\fam\slfam\csname\t@xt sl\@nd
\def\sc{\csname\smt@xt sl\@nd}}%
\textfont\bffam=\csname\t@xt bf\@nd   
\scriptfont\bffam=\csname\smt@xt bf\@nd
\scriptscriptfont\bffam=\csname\vst@xt bf\@nd
\def\bf{\fam\bffam\csname\t@xt bf\@nd
\def\sc{\csname\smt@xt bf\@nd}}%
\textfont\ttfam=\csname\t@xt tt\@nd   
\def\tt{\fam\ttfam\csname\t@xt tt\@nd}%
\textfont\bdifam=\csname\t@xt bdi\@nd 
\scriptfont\bdifam=\csname\smt@xt bdi\@nd
\scriptscriptfont\bdifam=\csname\vst@xt bdi\@nd
\def\bdi{\fam\bdifam\csname\t@xt bdi\@nd}%
\textfont\bsyfam=\csname\t@xt bsy\@nd 
\scriptfont\bsyfam=\csname\smt@xt bsy\@nd
\def\bsy{\fam\bsyfam\csname\t@xt bsy\@nd}%
\textfont\bssfam=\csname\t@xt bss\@nd 
\scriptfont\bssfam=\csname\smt@xt bss\@nd
\scriptscriptfont\bssfam=\csname\vst@xt bss\@nd
\def\bss{\fam\bssfam\csname\t@xt bss\@nd}%
\ifams
\textfont\msafam=\csname\t@xt msa\@nd 
\scriptfont\msafam=\csname\smt@xt msa\@nd
\scriptscriptfont\msafam=\csname\vst@xt msa\@nd
\textfont\msbfam=\csname\t@xt msb\@nd 
\scriptfont\msbfam=\csname\smt@xt msb\@nd
\scriptscriptfont\msbfam=\csname\vst@xt msb\@nd
\else\fi
\normalbaselineskip=\textb@seline pt
\setbox\strutbox=\hbox{\vrule height.7\normalbaselineskip 
depth.3\baselineskip width0pt}%
\def\sc{\csname\smt@xt rm\@nd}\normalbaselines\rm}
%
%
\def\smallfonts{%
\textfont0=\csname\sm@ll rm\@nd        
\scriptfont0=\csname\smsm@ll rm\@nd
\scriptscriptfont0=\csname\vssm@ll rm\@nd
\def\rm{\fam0\csname\sm@ll rm\@nd
\def\sc{\csname\smsm@ll rm\@nd}}%
\textfont1=\csname\sm@ll i\@nd         
\scriptfont1=\csname\smsm@ll i\@nd
\scriptscriptfont1=\csname\vssm@ll i\@nd
\textfont2=\csname\sm@ll sy\@nd        
\scriptfont2=\csname\smsm@ll sy\@nd
\scriptscriptfont2=\csname\vssm@ll sy\@nd
\textfont3=\csname\sm@ll ex\@nd        
\scriptfont3=\csname\smsm@ll ex\@nd
\scriptscriptfont3=\csname\smsm@ll ex\@nd
\textfont\itfam=\csname\sm@ll it\@nd   
\scriptfont\itfam=\csname\smsm@ll it\@nd
\scriptscriptfont\itfam=\csname\vssm@ll it\@nd
\def\it{\fam\itfam\csname\sm@ll it\@nd
\def\sc{\csname\smsm@ll it\@nd}}%
\textfont\slfam=\csname\sm@ll sl\@nd   
\def\sl{\fam\slfam\csname\sm@ll sl\@nd
\def\sc{\csname\smsm@ll sl\@nd}}%
\textfont\bffam=\csname\sm@ll bf\@nd   
\scriptfont\bffam=\csname\smsm@ll bf\@nd
\scriptscriptfont\bffam=\csname\vssm@ll bf\@nd
\def\bf{\fam\bffam\csname\sm@ll bf\@nd
\def\sc{\csname\smsm@ll bf\@nd}}%
\textfont\ttfam=\csname\sm@ll tt\@nd   
\def\tt{\fam\ttfam\csname\sm@ll tt\@nd}%
\textfont\bdifam=\csname\sm@ll bdi\@nd 
\scriptfont\bdifam=\csname\smsm@ll bdi\@nd
\scriptscriptfont\bdifam=\csname\vssm@ll bdi\@nd
\def\bdi{\fam\bdifam\csname\sm@ll bdi\@nd}%
\textfont\bsyfam=\csname\sm@ll bsy\@nd 
\scriptfont\bsyfam=\csname\smsm@ll bsy\@nd
\def\bsy{\fam\bsyfam\csname\sm@ll bsy\@nd}%
\textfont\bssfam=\csname\sm@ll bss\@nd 
\scriptfont\bssfam=\csname\smsm@ll bss\@nd
\scriptscriptfont\bssfam=\csname\vssm@ll bss\@nd
\def\bss{\fam\bssfam\csname\sm@ll bss\@nd}%
\ifams
\textfont\msafam=\csname\sm@ll msa\@nd 
\scriptfont\msafam=\csname\smsm@ll msa\@nd
\scriptscriptfont\msafam=\csname\vssm@ll msa\@nd
\textfont\msbfam=\csname\sm@ll msb\@nd 
\scriptfont\msbfam=\csname\smsm@ll msb\@nd
\scriptscriptfont\msbfam=\csname\vssm@ll msb\@nd
\else\fi
\normalbaselineskip=\smallb@seline pt%
\setbox\strutbox=\hbox{\vrule height.7\normalbaselineskip 
depth.3\baselineskip width0pt}%
\def\sc{\csname\smsm@ll rm\@nd}\normalbaselines\rm}%
%
 
%

%
\catcode`@=11
%
%

\catcode`\@=11
\def\vfootnote#1{\insert\footins\bgroup
\interlinepenalty=\interfootnotelinepenalty
\splittopskip=\ht\strutbox 
\splitmaxdepth=\dp\strutbox \floatingpenalty=20000
\leftskip=0pt \rightskip=0pt \spaceskip=0pt \xspaceskip=0pt%
\noindent\smallfonts\rm #1\ \ignorespaces\footstrut\futurelet\next\fo@t}
%
\def\hexnumber@#1{\ifcase#1 0\or 1\or 2\or 3\or 4\or 5\or 6\or 7\or 8\or
 9\or A\or B\or C\or D\or E\or F\fi}
\edef\bffam@{\hexnumber@\bffam}
\edef\bdifam@{\hexnumber@\bdifam}
\edef\bsyfam@{\hexnumber@\bsyfam}
\def\undefine#1{\let#1\undefined}
\def\newsymbol#1#2#3#4#5{\let\next@\relax
 \ifnum#2=\thr@@\let\next@\bdifam@\else
 \ifams
 \ifnum#2=\@ne\let\next@\msafam@\else
 \ifnum#2=\tw@\let\next@\msbfam@\fi\fi
 \fi\fi
 \mathchardef#1="#3\next@#4#5}
\def\mathhexbox@#1#2#3{\relax
 \ifmmode\mathpalette{}{\m@th\mathchar"#1#2#3}%
 \else\leavevmode\hbox{$\m@th\mathchar"#1#2#3$}\fi}
%

%
%
\ifams
%
%
\expandafter\chardef\csname pre amssym.def at\endcsname=\the\catcode`\@
\catcode`\@=11
\edef\msafam@{\hexnumber@\msafam}
\edef\msbfam@{\hexnumber@\msbfam}
\mathchardef\dabar@"0\msafam@39
\def\dashrightarrow{\mathrel{\dabar@\dabar@\mathchar"0\msafam@4B}}
\def\dashleftarrow{\mathrel{\mathchar"0\msafam@4C\dabar@\dabar@}}

\def\ulcorner{\delimiter"4\msafam@70\msafam@70 }
\def\urcorner{\delimiter"5\msafam@71\msafam@71 }
\def\llcorner{\delimiter"4\msafam@78\msafam@78 }
\def\lrcorner{\delimiter"5\msafam@79\msafam@79 }
\def\yen{{\mathhexbox@\msafam@55 }}
\def\checkmark{{\mathhexbox@\msafam@58 }}
\def\circledR{{\mathhexbox@\msafam@72 }}
\def\maltese{{\mathhexbox@\msafam@7A }}
\def\setboxz@h{\setbox\z@\hbox}
\def\wdz@{\wd\z@}
\def\widehat#1{\setboxz@h{$\m@th#1$}%
 \ifdim\wdz@>\tw@ em\mathaccent"0\msbfam@5B{#1}%
 \else\mathaccent"0362{#1}\fi}
\def\widetilde#1{\setboxz@h{$\m@th#1$}%
 \ifdim\wdz@>\tw@ em\mathaccent"0\msbfam@5D{#1}%
 \else\mathaccent"0365{#1}\fi}
%
\catcode`\@=\csname pre amssym.def at\endcsname
%
%

\expandafter\ifx\csname pre amssym.tex at\endcsname\relax \else  \fi
\expandafter\chardef\csname pre amssym.tex at\endcsname=\the\catcode`\@
\catcode`\@=11
%
\newsymbol\boxdot 1200
\newsymbol\boxplus 1201
\newsymbol\boxtimes 1202
\newsymbol\square 1003
\newsymbol\blacksquare 1004
\newsymbol\centerdot 1205
\newsymbol\lozenge 1006
\newsymbol\blacklozenge 1007
\newsymbol\circlearrowright 1308
\newsymbol\circlearrowleft 1309
\undefine\rightleftharpoons
\newsymbol\rightleftharpoons 130A
\newsymbol\leftrightharpoons 130B
\newsymbol\boxminus 120C
\newsymbol\Vdash 130D
\newsymbol\Vvdash 130E
\newsymbol\vDash 130F
\newsymbol\twoheadrightarrow 1310
\newsymbol\twoheadleftarrow 1311
\newsymbol\leftleftarrows 1312
\newsymbol\rightrightarrows 1313
\newsymbol\upuparrows 1314
\newsymbol\downdownarrows 1315
\newsymbol\upharpoonright 1316
 
\newsymbol\downharpoonright 1317
\newsymbol\upharpoonleft 1318
\newsymbol\downharpoonleft 1319
\newsymbol\rightarrowtail 131A
\newsymbol\leftarrowtail 131B
\newsymbol\leftrightarrows 131C
\newsymbol\rightleftarrows 131D
\newsymbol\Lsh 131E
\newsymbol\Rsh 131F
\newsymbol\rightsquigarrow 1320
\newsymbol\leftrightsquigarrow 1321
\newsymbol\looparrowleft 1322
\newsymbol\looparrowright 1323
\newsymbol\circeq 1324
\newsymbol\succsim 1325
\newsymbol\gtrsim 1326
\newsymbol\gtrapprox 1327
\newsymbol\multimap 1328
\newsymbol\therefore 1329
\newsymbol\because 132A
\newsymbol\doteqdot 132B
 
\newsymbol\triangleq 132C
\newsymbol\precsim 132D
\newsymbol\lesssim 132E
\newsymbol\lessapprox 132F
\newsymbol\eqslantless 1330
\newsymbol\eqslantgtr 1331
\newsymbol\curlyeqprec 1332
\newsymbol\curlyeqsucc 1333
\newsymbol\preccurlyeq 1334
\newsymbol\leqq 1335
\newsymbol\leqslant 1336
\newsymbol\lessgtr 1337
\newsymbol\backprime 1038
\newsymbol\risingdotseq 133A
\newsymbol\fallingdotseq 133B
\newsymbol\succcurlyeq 133C
\newsymbol\geqq 133D
\newsymbol\geqslant 133E
\newsymbol\gtrless 133F
\newsymbol\sqsubset 1340
\newsymbol\sqsupset 1341
\newsymbol\vartriangleright 1342
\newsymbol\vartriangleleft 1343
\newsymbol\trianglerighteq 1344
\newsymbol\trianglelefteq 1345
\newsymbol\bigstar 1046
\newsymbol\between 1347
\newsymbol\blacktriangledown 1048
\newsymbol\blacktriangleright 1349
\newsymbol\blacktriangleleft 134A
\newsymbol\vartriangle 134D
\newsymbol\blacktriangle 104E
\newsymbol\triangledown 104F
\newsymbol\eqcirc 1350
\newsymbol\lesseqgtr 1351
\newsymbol\gtreqless 1352
\newsymbol\lesseqqgtr 1353
\newsymbol\gtreqqless 1354
\newsymbol\Rrightarrow 1356
\newsymbol\Lleftarrow 1357
\newsymbol\veebar 1259
\newsymbol\barwedge 125A
\newsymbol\doublebarwedge 125B
\undefine\angle
\newsymbol\angle 105C
\newsymbol\measuredangle 105D
\newsymbol\sphericalangle 105E
\newsymbol\varpropto 135F
\newsymbol\smallsmile 1360
\newsymbol\smallfrown 1361
\newsymbol\Subset 1362
\newsymbol\Supset 1363
\newsymbol\Cup 1264
 
\newsymbol\Cap 1265
 
\newsymbol\curlywedge 1266
\newsymbol\curlyvee 1267
\newsymbol\leftthreetimes 1268
\newsymbol\rightthreetimes 1269
\newsymbol\subseteqq 136A
\newsymbol\supseteqq 136B
\newsymbol\bumpeq 136C
\newsymbol\Bumpeq 136D
\newsymbol\lll 136E
 
\newsymbol\ggg 136F
 
\newsymbol\circledS 1073
\newsymbol\pitchfork 1374
\newsymbol\dotplus 1275
\newsymbol\backsim 1376
\newsymbol\backsimeq 1377
\newsymbol\complement 107B
\newsymbol\intercal 127C
\newsymbol\circledcirc 127D
\newsymbol\circledast 127E
\newsymbol\circleddash 127F
\newsymbol\lvertneqq 2300
\newsymbol\gvertneqq 2301
\newsymbol\nleq 2302
\newsymbol\ngeq 2303
\newsymbol\nless 2304
\newsymbol\ngtr 2305
\newsymbol\nprec 2306
\newsymbol\nsucc 2307
\newsymbol\lneqq 2308
\newsymbol\gneqq 2309
\newsymbol\nleqslant 230A
\newsymbol\ngeqslant 230B
\newsymbol\lneq 230C
\newsymbol\gneq 230D
\newsymbol\npreceq 230E
\newsymbol\nsucceq 230F
\newsymbol\precnsim 2310
\newsymbol\succnsim 2311
\newsymbol\lnsim 2312
\newsymbol\gnsim 2313
\newsymbol\nleqq 2314
\newsymbol\ngeqq 2315
\newsymbol\precneqq 2316
\newsymbol\succneqq 2317
\newsymbol\precnapprox 2318
\newsymbol\succnapprox 2319
\newsymbol\lnapprox 231A
\newsymbol\gnapprox 231B
\newsymbol\nsim 231C
\newsymbol\ncong 231D
\newsymbol\diagup 231E
\newsymbol\diagdown 231F
\newsymbol\varsubsetneq 2320
\newsymbol\varsupsetneq 2321
\newsymbol\nsubseteqq 2322
\newsymbol\nsupseteqq 2323
\newsymbol\subsetneqq 2324
\newsymbol\supsetneqq 2325
\newsymbol\varsubsetneqq 2326
\newsymbol\varsupsetneqq 2327
\newsymbol\subsetneq 2328
\newsymbol\supsetneq 2329
\newsymbol\nsubseteq 232A
\newsymbol\nsupseteq 232B
\newsymbol\nparallel 232C
\newsymbol\nmid 232D
\newsymbol\nshortmid 232E
\newsymbol\nshortparallel 232F
\newsymbol\nvdash 2330
\newsymbol\nVdash 2331
\newsymbol\nvDash 2332
\newsymbol\nVDash 2333
\newsymbol\ntrianglerighteq 2334
\newsymbol\ntrianglelefteq 2335
\newsymbol\ntriangleleft 2336
\newsymbol\ntriangleright 2337
\newsymbol\nleftarrow 2338
\newsymbol\nrightarrow 2339
\newsymbol\nLeftarrow 233A
\newsymbol\nRightarrow 233B
\newsymbol\nLeftrightarrow 233C
\newsymbol\nleftrightarrow 233D
\newsymbol\divideontimes 223E
\newsymbol\varnothing 203F
\newsymbol\nexists 2040
\newsymbol\Finv 2060
\newsymbol\Game 2061
\newsymbol\mho 2066
\newsymbol\eth 2067
\newsymbol\eqsim 2368
\newsymbol\beth 2069
\newsymbol\gimel 206A
\newsymbol\daleth 206B
\newsymbol\lessdot 236C
\newsymbol\gtrdot 236D
\newsymbol\ltimes 226E
\newsymbol\rtimes 226F
\newsymbol\shortmid 2370
\newsymbol\shortparallel 2371
\newsymbol\smallsetminus 2272
\newsymbol\thicksim 2373
\newsymbol\thickapprox 2374
\newsymbol\approxeq 2375
\newsymbol\succapprox 2376
\newsymbol\precapprox 2377
\newsymbol\curvearrowleft 2378
\newsymbol\curvearrowright 2379
\newsymbol\digamma 207A
\newsymbol\varkappa 207B
\newsymbol\Bbbk 207C
\newsymbol\hslash 207D
\undefine\hbar
\newsymbol\hbar 207E
\newsymbol\backepsilon 237F
%
%
\catcode`\@=\csname pre amssym.tex at\endcsname

\fi
%
%
\newsymbol\bitGamma 3000
\newsymbol\bitDelta 3001
\newsymbol\bitTheta 3002
\newsymbol\bitLambda 3003
\newsymbol\bitXi 3004
\newsymbol\bitPi 3005
\newsymbol\bitSigma 3006
\newsymbol\bitUpsilon 3007
\newsymbol\bitPhi 3008
\newsymbol\bitPsi 3009
\newsymbol\bitOmega 300A
\newsymbol\balpha 300B
\newsymbol\bbeta 300C
\newsymbol\bgamma 300D
\newsymbol\bdelta 300E
\newsymbol\bepsilon 300F
\newsymbol\bzeta 3010
\newsymbol\bfeta 3011
\newsymbol\btheta 3012
\newsymbol\biota 3013
\newsymbol\bkappa 3014
\newsymbol\blambda 3015
\newsymbol\bmu 3016
\newsymbol\bnu 3017
\newsymbol\bxi 3018
\newsymbol\bpi 3019
\newsymbol\brho 301A
\newsymbol\bsigma 301B
\newsymbol\btau 301C
\newsymbol\bupsilon 301D
\newsymbol\bphi 301E
\newsymbol\bchi 301F
\newsymbol\bpsi 3020
\newsymbol\bomega 3021
\newsymbol\bvarepsilon 3022
\newsymbol\bvartheta 3023
\newsymbol\bvaromega 3024
\newsymbol\bvarrho 3025
\newsymbol\bvarzeta 3026
\newsymbol\bvarphi 3027
\newsymbol\bpartial 3040
\newsymbol\bell 3060
\newsymbol\bimath 307B
\newsymbol\bjmath 307C
\mathchardef\bnabla "0\bsyfam@72
\mathchardef\bdot "2\bsyfam@01
\mathchardef\bGamma "0\bffam@00
\mathchardef\bDelta "0\bffam@01
\mathchardef\bTheta "0\bffam@02
\mathchardef\bLambda "0\bffam@03
\mathchardef\bXi "0\bffam@04
\mathchardef\bPi "0\bffam@05
\mathchardef\bSigma "0\bffam@06
\mathchardef\bUpsilon "0\bffam@07
\mathchardef\bPhi "0\bffam@08
\mathchardef\bPsi "0\bffam@09
\mathchardef\bOmega "0\bffam@0A
\mathchardef\itGamma "0100
\mathchardef\itDelta "0101
\mathchardef\itTheta "0102
\mathchardef\itLambda "0103
\mathchardef\itXi "0104
\mathchardef\itPi "0105
\mathchardef\itSigma "0106
\mathchardef\itUpsilon "0107
\mathchardef\itPhi "0108
\mathchardef\itPsi "0109
\mathchardef\itOmega "010A
\mathchardef\Gamma "0000
\mathchardef\Delta "0001
\mathchardef\Theta "0002
\mathchardef\Lambda "0003
\mathchardef\Xi "0004
\mathchardef\Pi "0005
\mathchardef\Sigma "0006
\mathchardef\Upsilon "0007
\mathchardef\Phi "0008
\mathchardef\Psi "0009
\mathchardef\Omega "000A
%
%
\def\c@ps#1\endc@ps{\uppercase{\def\C@PS{#1}}}
%
%
%
\newif\ifblank
\newif\ifsided
\newif\iftwofigs
\newif\ifwidetab
%
%
\newbox\captionbox
\newbox\tablebox
\newbox\numberb@x
%
%
\newdimen\columnwidth
\newdimen\columnsep
\newdimen\pagewidth
\newdimen\pagedepth
\newdimen\doublepage
\newdimen\normalindent
\newdimen\headindent
\newdimen\leftindent
\newdimen\rightindent
\newdimen\floatsize
\newdimen\figsp@ce
\newdimen\tablewidth
\newdimen\alpharefindent
\newdimen\numrefindent
\newdimen\tempindent
\newdimen\digitwidth    
%
%
\newskip\abovetitleskip
\newskip\belowtitleskip
\newskip\abovetypeskip
\newskip\belowtypeskip
\newskip\belowauthorskip
\newskip\belowaddressskip
\newskip\aboveabstractskip
\newskip\abovesecskip
\newskip\belowsecskip
\newskip\smallsecskip
\newskip\abovesubskip
\newskip\belowsubskip
\newskip\smallsubskip
\newskip\abovesubsubskip
\newskip\belowfigskip
\newskip\abovetableskip
\newskip\belowtableskip
\newskip\noinsertskip
%
%
\newcount\firstp@ge
\newcount\secno      
\newcount\subno      
\newcount\subsubno   
\newcount\appno      
\newcount\tabno      
\newcount\tabnum     
\newcount\figno      
\newcount\fignum     
\newcount\countno    
\countno=1
\def\nextfig{\chapnum.\the\fignum}
\def\nexttab{\chapnum.\the\tabnum}
\def\lastfig{\chapnum.\the\figno}
\def\lasttab{\chapnum.\the\tabno}
%
%
\def\gac{\global\advance\countno by 1}
\def\en{\eqno(\the\countno)\gac}
\def\aen{&(\the\countno)\gac}
\def\enpt#1{\eqno(\the\countno#1)}
%
%


%

\def\etal{{\it et al\/}\ }
\def\frac#1#2{{#1\over#2}}

\ifams
\gdef\lap{\lesssim}
\gdef\gap{\gtrsim}

\else
\def\gap{\;\lower3pt\hbox{$\buildrel > \over \sim$}\;}%
\def\lap{\;\lower3pt\hbox{$\buildrel < \over \sim$}\;}\fi

\chardef\ii="10
\def\tqs{\hbox to 25pt{\hfil}}


\def\pt(#1){({\it #1\/})}
%

%
%
\def\rp{\raise8pt\hbox{$\scriptstyle\prime$}}
%
%

%

%
%
\def\slashchar#1{\setbox0=\hbox{$#1$}\dimen0=\wd0%
\setbox1=\hbox{/}\dimen1=\wd1%
\ifdim\dimen0>\dimen1%
\rlap{\hbox to \dimen0{\hfil/\hfil}}#1\else                                        
\rlap{\hbox to \dimen1{\hfil$#1$\hfil}}/\fi}
%
%
\def\textindent#1{\noindent\hbox to \parindent{#1\hss}\ignorespaces}
%
%
%
\def\ragged{\rightskip=0pt plus8em
    \hyphenpenalty=10000\exhyphenpenalty=10000}
\def\nowtext{\noindent\textfonts\rm\ignorespaces}
%
%
\def\opencirc{\raise1pt\hbox{$\scriptstyle{\bigcirc}$}}

\ifams
\def\opensqr{\hbox{$\square$}}

\def\opentridown{\hbox{$\triangledown$}}

\else
\def\opensqr{\vbox{\hrule height.4pt\hbox{\vrule width.4pt height3.5pt
    \kern3.5pt\vrule width.4pt}\hrule height.4pt}}

\def\opentridown{\raise1pt\hbox{$\scriptstyle\bigtriangledown$}}

\fi

%
%

%
%
\def\m@th{\mathsurround=0pt}
\def\cases#1{%
\left\{\,\vcenter{\normalbaselines\openup1\jot\m@th%
     \ialign{$##\hfil$&\rm\tqs##\hfil\crcr#1\crcr}}\right.}%
%
%
\def\oldcases#1{\left\{\,\vcenter{\normalbaselines\m@th
    \ialign{$##\hfil$&\rm\quad##\hfil\crcr#1\crcr}}\right.}
\def\leavemode{\unhbox\voidb@x}
\def\AA{\leavemode\setbox0=\hbox{h}\dimen@=\ht0 \advance\dimen@ by-1ex
 \rlap{\raise.67\dimen@\hbox{\char'27}}{\rm A}}
%
%
\overfullrule=0pt
%
%

%
%
%

%
%

%
%

%
%

%
\def\lineup{\setbox0=\hbox{\smallfonts\rm 0}%
    \digitwidth=\wd0%
    \def\0{\kern\digitwidth}
    \def\\{\hbox{$\phantom{-}$}}
    \def\-{\llap{$-$}}}
%
%



%
%

%
\catcode`@=12
%
 
%

%
%
\headsize{17}{20}
\textsize{12}{14}
\smallsize{10}{12}
\catcode`@=11
\hsize=37.5pc
\pagewidth=37.5pc
\columnwidth=18pc
\columnsep=\pagewidth
\global\advance\columnsep by -2\columnwidth
\vsize=57.5pc
\pagedepth=57.5pc
\doublepage=116pc
\lineskip=0pt
\parskip=0pt plus0.1pt
\hfuzz=1pt  
\vfuzz=2pt
\tolerance=5000
\widowpenalty=1000
\clubpenalty=500
\brokenpenalty=500
\predisplaypenalty=100
\nopagenumbers
\voffset=0pt
\def\page#1{\global\pageno=#1\global\firstp@ge=\pageno}
\page{1}
%
%
%
\parindent=18pt
\normalindent=18pt
%
%
\headindent=5pc
%
%
\leftindent=3pc
\rightindent=3pc
%
%
\alpharefindent=20pt
\numrefindent=42pt
%
%
\abovetitleskip=28pt plus7pt minus14pt
\belowtitleskip=21pt plus7pt minus 7pt
\abovetypeskip=-7pt plus 7pt
\belowtypeskip=-7pt plus 7pt
\def\abovetitlespace{\vglue\abovetitleskip\relax}
\def\belowtitlespace{\vskip\belowtitleskip\relax}
\def\abovetypespace{\vglue\abovetypeskip\relax}
\def\belowtypespace{\vskip\belowtypeskip\relax}
%
%
\belowauthorskip=21pt plus7pt minus7pt
\belowaddressskip=7pt plus 7pt
\def\belowauthorspace{\vskip\belowauthorskip\relax}
\def\belowaddressspace{\vskip\belowaddressskip\relax}
\aboveabstractskip=28pt plus14pt minus14pt
\def\aboveabstractspace{\vskip\aboveabstractskip\relax}
%
%
\abovesecskip=28pt plus7pt minus14pt
\belowsecskip=14pt
\smallsecskip=7pt plus 7pt
\def\abovesecspace{\vskip\abovesecskip\relax}
\def\belowsecspace{\vskip\belowsecskip\relax}
\def\smallsecspace{\vskip-\smallsecskip\relax}
%
%
\abovesubskip=21pt plus7pt minus7pt
\belowsubskip=7pt
\smallsubskip=7pt plus 7pt
\def\abovesubspace{\vskip\abovesecskip\relax}
\def\belowsubspace{\vskip\belowsecskip\relax}
\def\smallsubspace{\vskip-\smallsecskip\relax}
%
%
\abovesubsubskip=14pt plus10pt minus4pt
\def\abovesubsubspace{\vskip\abovesubsubskip\relax}
%
%
\belowfigskip=1pc plus3pt minus2pt
\def\belowfigspace{\vskip\belowfigskip\relax}
%
%
\floatsize=\pagewidth
\global\advance\floatsize by -\leftindent
\global\advance\floatsize by -\rightindent
\noinsertskip=21pt plus7pt minus7pt
\def\noinsertspace{\vskip\noinsertskip\relax}
%
%
\abovetableskip=3pt
\belowtableskip=0pt
%
%
\noinsertskip=24pt plus 6pt minus 6pt
\def\noinsertspace{\vskip\noinsertskip\relax}
%
%
\headline={\ifnum\pageno=\firstp@ge\hfill\else
            \ifodd\pageno\righthead
                   \else\lefthead\fi\fi}
\def\righthead{\hfill\textfonts\it\papertitle\hfill
    \llap{\textfonts\rm\folio}}
\def\lefthead{\rlap{\textfonts
    \rm\folio}\hfill\textfonts\it\c@nftitle\hfill}
\def\conftitle#1{\gdef\c@nftitle{#1}}
\conftitle{}
%
%
%
%
\def\title#1{\abovetitlespace
   {\noindent\headfonts\bf\ragged#1\par}%
    \gdef\papertitle{#1}\futurelet\next\sh@rt} 
\def\sh@rt{\ifx\next[\let\next=\sh@rttitle\else
                     \let\next=\belowtitlespace\fi\next}
\def\sh@rttitle[#1]{\gdef\papertitle{#1}\belowtitlespace}
%
%
\def\review#1{\abovetypespace
    {\parindent=\headindent
     \textfonts\bf REVIEW\par}%
     \belowtypespace
    {\noindent\headfonts\bf\ragged#1\par}%
     \gdef\papertitle{#1}\futurelet\next\sh@rt} 
%
%
\def\invited#1{\abovetypespace
    {\parindent=\headindent
     \textfonts\bf INVITED PAPER\par}%
     \belowtypespace
    {\noindent\headfonts\bf\ragged#1\par}%
     \gdef\papertitle{#1}\futurelet\next\sh@rt} 
%
%
\def\author#1{{\parindent=\headindent\hang\textfonts\bf\ragged#1\par}%
    \belowauthorspace}
%
%
\def\address#1{{\parindent=\headindent\hang\textfonts\rm\ragged#1\par}%
    \belowaddressspace}
%
%
\def\beginabstract{\aboveabstractspace
     \parindent=\headindent\hang\textfonts{\bf Abstract. }\rm}
\def\endabstract{\par\parindent=\normalindent\textfonts\rm}
%
%

%
%
%
%
\def\section#1{\goodbreak\abovesecspace\nobreak
    \subno=0\subsubno=0\global\advance\secno by 1%
   {\noindent\textfonts\bf\the\secno. \ragged#1\par}%
    \nobreak\futurelet\next\s@b}
\def\s@b{\ifx\next*\let\next=\smallsecsp@ce\else
                   \let\next=\bigsecsp@ce\fi\next}
\def\smallsecsp@ce#1{\smallsecspace\nobreak\nowtext}
\def\bigsecsp@ce{\belowsecspace\nobreak\nowtext}
%
%
\def\subsection#1{\goodbreak\abovesubspace\nobreak
     \subsubno=0\global\advance\subno by 1
    {\noindent\textfonts\it\the\secno.\the\subno. \ragged#1\par}%
     \nobreak\futurelet\next\subs@b}
\def\subs@b{\ifx\next*\let\next=\smallsubsp@ce\else
                      \let\next=\bigsubsp@ce\fi\next}
\def\smallsubsp@ce#1{\smallsubspace\nobreak\nowtext}
\def\bigsubsp@ce{\belowsubspace\nobreak\nowtext}
%
%
\def\subsubsection#1{\goodbreak\abovesubsubspace\nobreak
     \global\advance\subsubno by 1
    {\noindent\textfonts\it\the\secno.\the\subno.\the\subsubno.
     #1\null.}\quad\ignorespaces}
%
%
%
\def\numappendix#1{\goodbreak\abovesecspace\nobreak
    \subno=0\subsubno=0\global\advance\appno by 1
   {\noindent\textfonts\bf Appendix \the\appno. \ragged#1\par}%
    \nobreak\futurelet\next\s@b}
%
%
\def\numsubappendix#1{\goodbreak\abovesubspace\nobreak
    \subsubno=0\global\advance\subno by 1
   {\noindent\textfonts\it A\the\appno.\the\subno. #1\par}%
    \nobreak\futurelet\next\subs@b}
%
%
%
\def\Appendix#1{\goodbreak\abovesecspace\nobreak
    \subno=0\subsubno=0
   {\noindent\textfonts\bf Appendix. #1\par}%
    \nobreak\futurelet\next\s@b}
%
%
\def\appendix#1#2{\goodbreak\abovesecspace\nobreak
    \subno=0\subsubno=0
   {\noindent\textfonts\bf Appendix #1. \ragged #2\par}%
    \nobreak\futurelet\next\s@b}
%
%

%
%
\def\subappendix#1#2{\goodbreak\abovesubspace\nobreak
    \subsubno=0\global\advance\subno by 1
   {\noindent\textfonts\it #1\the\subno. \ragged#2\par}%
    \nobreak\futurelet\next\subs@b}
%
%
\def\ack{\goodbreak\abovesecspace\nobreak
   {\noindent\textfonts\bf Acknowledgments\par}%
    \nobreak\belowsecspace\nobreak\nowtext}
%

%
%
%
%
%
%
\def\figure#1#2{\fignumber\figsp@ce=#1
    \def\figc@p{#2}\futurelet\next\figpl@cement}
\def\figpl@cement{\ifx\next[\let\next=\nsfigpl@ce
                  \else\let\next=\figpl@ce\fi\next}
\def\figpl@ce{\let\instype=\topinsert\ifsided\sidedc@ption
    \else\c@ption\fi}
\def\c@ption{\instype\ifx\instype\pageinsert\vfill
    \else\vskip\figsp@ce\belowfigspace\fi
    \iftwofigs\doublec@ption\else\figcaption\fi\closefig}
\def\nsfigpl@ce[#1]{\def\@pos{#1}%
                    \if\@pos m\let\instype=\midinsert\else
                    \if\@pos b\let\instype=\botinsert\else
                    \if\@pos p\let\instype=\pageinsert\else
                    \if\@pos h\let\instype=\noinsert\else
                       \let\instype=\topinsert\fi\fi\fi\fi
    \c@ptiontype}
\def\c@ptiontype{\ifsided\sidedc@ption
                         \else\c@ption\fi}
\def\botinsert{\insert\footins\bgroup}
\def\noinsert{\noinsertspace\vbox\bgroup}
\def\fignumber{\global\advance\figno by 1
    \global\advance\fignum by 1}
\def\closefig{\ifx\instype\botinsert\egroup\else
              \ifx\instype\noinsert\egroup\noinsertspace
                  \else\endinsert\fi\fi}
\def\pagefigure#1{\fignumber\def\figc@p{#1}%
    \let\instype=\pageinsert\c@ption}
\def\figblank{\pageinsert\global\advance\figno by 1\vfill\endinsert}
\def\figcaption{\setbox\captionbox=\hbox{\smallfonts
    \bf Figure \the\figno.\quad\rm \figc@p}%
    \ifdim\wd\captionbox>\floatsize\leftskip=\leftindent
          \rightskip=\rightindent{\noindent\smallfonts\rm
          \unhbox\captionbox\par}%
          \leftskip=0pt\rightskip=0pt\else
          \centerline{\unhbox\captionbox}\fi}
\def\sidedcaption#1#2#3{\global\sidedtrue
    \fignumber\figsp@ce=#1\tempindent=#2
    \def\figc@p{#3}\futurelet\next\figpl@cement}
\def\sidedc@ption{\instype\ifx\instype\pageinsert
    \vbox to\pagedepth\bgroup
    \else\vbox to\figsp@ce\bgroup\fi\vfill
    \sidedc@p\global\sidedfalse}
\def\sidedc@p{{\parindent=\tempindent
    \hang\smallfonts\rm{\bf Figure \the\figno.}\quad\figc@p
    \par}\egroup\closefig}
\def\side#1#2#3{\fignumber\figsp@ce=#1\global\twofigstrue
    \def\figc@p{#2}\def\c@ptwo{#3}%
    \futurelet\next\figpl@cement}

\def\doublec@ption{\noindent\hbox{\vtop{\hsize=\columnwidth
    \noindent\smallfonts\rm{\bf Figure \the\figno.}\quad
    \rm\figc@p\par}\hskip\columnsep
    \vtop{\hsize=\columnwidth
    \noindent\fignumber
    \smallfonts\rm{\bf Figure \the\figno.}\quad\rm\c@ptwo
    \par}}\par\global\twofigsfalse} 
%
%
\def\tabnumber{\widetabfalse\global\advance\tabno by 1
    \global\advance\tabnum by 1}
\def\table#1{\gdef\tabc@p{#1}\tabnumber
    \futurelet\next\tabpl@cement}
\def\tablecont{\gdef\tabc@p{(continued)}\widetabfalse
    \futurelet\next\tabpl@cement}%
\def\tabpl@cement{\ifx\next*\let\instype=\topinsert
                            \let\next=\widetablec@p\else
                  \ifx\next[\let\next=\nstabpl@ce
                      \else\let\instype=\topinsert
                           \let\next=\tablec@ption\fi\fi
                      \next}
\def\nstabpl@ce[#1]{\def\@pos{#1}%
                    \if\@pos m\let\instype=\midinsert\else
                    \if\@pos b\let\instype=\botinsert\else
                    \if\@pos p\let\instype=\pageinsert\else
                    \if\@pos h\let\instype=\noinsert\else
   \let\instype=\topinsert\fi\fi\fi\fi
     \futurelet\nextok\tablec@psize}
\def\tablec@psize{\ifx\nextok*\let\nextok=\widetablec@p
                 \else\let\nextok=\tablec@ption\fi\nextok} 
\def\widetablec@p#1{\global\widetabtrue\instype
    \vbox\bgroup\t@bcaption}
\def\tablec@ption{\instype\vbox\bgroup\t@bcaption}
\def\t@bcaption{\setbox\captionbox=\hbox{\smallfonts
    \bf Table \the\tabno.\quad\rm \tabc@p}%
    \ifdim\wd\captionbox>\floatsize\leftskip=\leftindent
          \rightskip=\rightindent{\noindent\smallfonts\rm
          \unhbox\captionbox\par}%
          \leftskip=0pt\rightskip=0pt\else
          \centerline{\unhbox\captionbox}\fi}
\def\align{\abovedisplayskip=0pt\belowdisplayskip=0pt%
    $$\vbox\bgroup\smallfonts\rm\tabskip=0pt%
    \setbox\tablebox=\vbox\bgroup
    \halign\ifwidetab to \pagewidth\fi
    \bgroup\strut\tabskip=1.2pc plus1pc minus .5pc}
\def\endalign{\egroup\egroup\global\tablewidth=\wd\tablebox
    \unvbox\tablebox\egroup$$}
\def\endtable{\egroup\ifx\instype\botinsert\egroup\else
              \ifx\instype\noinsert\egroup\noinsertspace\else
              \ifx\instype\pageinsert\vfill\fi
                  \endinsert\fi\fi}
\def\pagetable#1{\tabnumber\def\tabc@p{#1}%
    \let\instype=\pageinsert\futurelet\nextok\tablec@psize}
\def\tabblank{\pageinsert\global\advance\tabno by 1\vfill\endinsert}
%
%
\def\references{\goodbreak\abovesecspace\nobreak
   {\noindent\textfonts\bf References\par}%
    \alpharefs\nobreak\bigsecsp@ce}
\def\alpharefs{\tempindent=\parindent\parindent=0pt
    \everypar{\hangindent=\alpharefindent
    \hangafter=1\frenchspacing\rm}}
\def\numrefs#1{\setbox\numberb@x=\hbox{\textfonts\rm[#1]\quad}%
   \everypar{\parindent=\wd\numberb@x\hangindent=\numrefindent
   \hangafter=1\frenchspacing\rm}}
\def\ref#1{\par\noindent\hbox to\parindent{\hss[#1]\quad}%
   \ignorespaces}
\def\nonum{\par\noindent\hbox to\parindent{\hfill}\ignorespaces}
\def\endrefs{\everypar{}\parindent=\tempindent\textfonts\rm}
\catcode`@=12

{\obeylines\gdef\startdisplay#1
  {\catcode`\^^M=5$$#1\halign\bgroup\indent##\hfil&&\qquad##\hfil\cr}}
\outer\def\enddisplay{\crcr\egroup$$}

\chardef\other=12
\def\ttverbatim{\begingroup \catcode`\\=\other \catcode`\{=\other
  \catcode`\}=\other \catcode`\$=\other \catcode`\&=\other
  \catcode`\#=\other \catcode`\%=\other \catcode`\~=\other
  \catcode`\_=\other \catcode`\^=\other
  \obeyspaces \obeylines \tt}
{\obeyspaces\gdef {\ }}  

\outer\def\begintt{$$\let\par=\endgraf \ttverbatim \parskip=0pt
  \catcode`\|=0 \rightskip=-5pc \ttfinish}
{\catcode`\|=0 |catcode`|\=\other 
  |obeylines 
  |gdef|ttfinish#1^^M#2\endtt{#1|vbox{#2}|endgroup$$}}

\catcode`\|=\active
{\obeylines\gdef|{\ttverbatim\spaceskip=.5em plus.25em minus.15em\let^^M=\ \let|=\endgroup}}%

\pretolerance=10000 
%
%
\abovedisplayskip=0pt plus6pt
\belowdisplayskip=0pt plus6pt
\abovedisplayshortskip=-6pt plus 6pt
\belowdisplayshortskip=0pt plus6pt
%

\def\ApJstyle{\scriptstyle}
\def\etal{et~al.\ }
\def\gapprox{\hbox{\lower .8ex\hbox{$\,\buildrel > \over\sim\,$}}}
\def\lapprox{\hbox{\lower .8ex\hbox{$\,\buildrel < \over\sim\,$}}}
\newcount\counttemp
\newcount\counteqn
\counteqn=0
\def\eqnnumber#1){\number\counteqn \rm #1)}
\def\eqn#1){\global\advance\counteqn by 1   \eqnnumber \rm #1)}
\def\aeqn#1){\global\advance\counteqn by 1   \eqnnumber \rm #1)}
\def\showeqn#1){\counttemp=\counteqn \advance\counttemp by #1
                \number\counttemp)}
\def\showeqnsq#1]{\counttemp=\counteqn \advance\counttemp by #1
                \number\counttemp]}
\def\showeqnx#1,#2){\counttemp=\counteqn \advance\counttemp by #1
                \number\counttemp#2)}

\conftitle{Second Oak Ridge symposium on atomic \& nuclear astrophysics}

\vglue -5.\baselineskip       

\title{A grey $\gamma$-ray transfer procedure for supernovae} 

\author{David J Jeffery}

\belowaddressskip=3pt plus 3pt        

\address{Physics Division, Oak Ridge National Laboratory,
Oak Ridge, Tennessee 37831-6372, U.S.A.;  jeffery@mail.phy.ornl.gov;
http://www-cfadc.phy.ornl.gov/}

\address{To appear 1998 in {\it proc.~second Oak Ridge symposium on atomic
         \& nuclear astrophysics} ed~A~Mezzacappa (Bristol:  Institute
         of Physics Publ.), astro-ph/9802229}  


\beginabstract
The $\gamma$-ray transfer in supernovae for the purposes
of energy deposition in the ejecta can be approximated
as grey radiative transfer using
mean opacities.
In past work there is a single pure absorption mean opacity
which is a free parameter.
Accurate results can be obtained by varying this mean opacity
to fit the results of more accurate procedures. 
In this paper, we present a grey $\gamma$-ray transfer
procedure for energy deposition in which there are multiple
mean opacities that are not free parameters and that have
both absorption and scattering components.
This procedure is based on a local-state (LS) approximation, and
so we call it the 
LS grey $\gamma$-ray transfer
procedure or LS~procedure for short.
\endabstract

\section{Introduction}
A major source of observable supernova luminosity, and in
the case of Type~Ia supernovae (SNe~Ia) virtually the only source,
is the decay energy of radioactive species synthesized in the
explosion.
The overwhelmingly dominant decay chain for the observable
period of most supernovae is
$^{\ApJstyle 56}$$\rm Ni
\to^{\ApJstyle 56}$$\rm Co\to^{\ApJstyle 56}$$\rm Fe$
with \hbox{half-lives} of 5.9 and~77.27~days for the first and second
decays, respectively (Huo 1992).
The first decay releases almost all its energy in the form of
$\gamma$-rays and the second in $\gamma$-rays and, in $19\,$\% of
the decays, in positrons (Browne and~Firestone 1986;  Huo 1992).
Until very late times almost all the positrons deposit their kinetic
energy in the ejecta and annihilate to form more $\gamma$-rays.

     The $\gamma$-ray and positron kinetic energy deposited in the
ejecta is needed to calculate the supernova
ultraviolet-optical-infrared luminosity.
This deposited energy is effectively in the form of fast
electrons.
The usual assumption (which may not be entirely correct
[e.g.,  Colgate~\etal 1980;  Chan and~Lingenfelter
1993;  Ruiz-Lapuente 1997;  Milne~\etal 1997]) is that
the positron kinetic energy is deposited locally.
For the $\gamma$-ray deposition, a $\gamma$-ray
transfer procedure is needed.
A full treatment of the $\gamma$-ray transfer requires computationally
intensive procedures such as Monte Carlos.
But if one just requires the energy deposited in the form
of fast electrons without knowing the fast electron spectrum,
then a simple, time-independent, static,
grey (i.e., frequency-integrated)
radiative transfer procedure using
a single pure absorption mean opacity can be quite accurate
as shown by
Colgate~\etal (1980), Sutherland and~Wheeler (1984),
Ambwani and~Sutherland (1988), and
Swartz~\etal (1995, hereafter SSH).
(Note there is no practical mean opacity that will
reduce $\gamma$-ray transfer in supernovae to exact
grey radiative transfer.
Thus grey supernova $\gamma$-ray transfer procedures will
always be approximations.)

     The mean opacity used in past work is a parameter
chosen by the comparison of the grey $\gamma$-ray transfer
results to those of more accurate $\gamma$-ray transfer
procedures.
Only one value being used for all the ejecta.
In the particular procedure of SSH, the mean opacity is
explicitly varied to obtain optimum accuracy for each
supernova epoch by comparison to
Monte Carlo results.
For the supernova model examined by SSH (SN~Ia model~W7
[Thielemann~\etal 1986]), the optimized SSH~procedure
obtained an accuracy
of a few percent locally  
and $2\,$\% globally.     
Despite the high accuracy achievable, the SSH procedure
is not completely satisfactory because the optimization requires
having done the full $\gamma$-ray transfer one wishes
to avoid by using a grey $\gamma$-ray transfer
procedure.

     In this paper we briefly present a time-independent, static,
grey $\gamma$-ray transfer procedure with no free parameters
that we believe will be
generally accurate at least globally to within a few percent.
This procedure is based on a local-state (LS) approximation, and
so we call it the LS grey radiative transfer
procedure or LS~procedure for short.
A full presentation of the LS~procedure is given by
Jeffery (1998, hereafter J98).

     In \S~2 we discuss the opacities and mean opacities
that are used in the LS~procedure.
Section~3 introduces the LS~procedure itself.
Conclusions are given in \S~4.

\section{Opacities}
In the energy range $0.05$--$50\,$MeV, $\gamma$-rays interact with
matter principally through three processes:
(1)~pair production in the Coulomb field of a nucleus or an electron,
(2)~the
photoelectric effect with bound electrons (which is just $\gamma$-ray
photoionization of an atom or ion), and
(3)~Compton scattering off electrons (e.g., Davisson 1965, p.~37).
Almost all the $\gamma$-rays from the
$^{\ApJstyle 56}$$\rm Ni
\to^{\ApJstyle 56}$$\rm Co\to^{\ApJstyle 56}$$\rm Fe$
decay chain and other decay chains important in supernovae
lie in the energy range $0.05$--$50\,$MeV and no $\gamma$-rays
exceed $\sim 3.6\,$MeV in fact (Browne and~Firestone 1986;  Huo 1992).
Thus the three mentioned processes determine $\gamma$-ray opacity in
supernovae (see also SSH's Fig.~1).
Of the three opacities Compton opacity is dominant:
in solar composition overwhelmingly so;
in metal-rich composition (such as in SNe~Ia) photoelectric opacity
(a pure absorption opacity) is also important (e.g., J98).

     Because of the dominance of Compton opacity, one expects
scattered $\gamma$-ray fields (viz., 1st, 2nd, 3rd, etc.~order
scattered fields) to be of some importance alongside
the initial (i.e., 0th order scattered) $\gamma$-ray field from
the radioactive decays and positron annihilation.
%
%
Nevertheless earlier grey $\gamma$-ray transfer procedures use a pure
absorption mean opacity.
The reason this works well is that for $\gamma$-rays of
energy typical of the radioactive decays in supernovae 
a large component of the Compton opacity
is nearly-forward, nearly-coherent scattering.
From a radiative transfer point of view, this component
is effectively almost neglectable.
The non-neglectable component of Compton scattering
is largely an absorption opacity.
By using a pure absorption mean opacity the earlier
procedures effectively neglected the nearly-forward,
nearly-coherent scattering.
This mean opacity incorporated all the absorption opacity
encountered by the 0th order field plus an extra absorption
component that accounted for the absorption of
the relatively weak scattered fields arising from the 0th order
non-forward, noncoherent scattering.
This extra absorption component is the actual underlying free
parameter of the SSH procedure.

     For the LS~procedure we choose to deal with the scattered fields
explicitly albeit crudely.
First, we will approximate the 0th order $\gamma$-ray field as a
line spectrum.
(The decay $\gamma$-ray spectrum is
almost exactly a line spectrum and
we assume that the positron annihilation can be approximated
as always going by the
two-$m_{\ApJstyle e}c^{\ApJstyle 2}$-photon process.)
Second, we approximate the Compton opacity by subtracting the 
nearly-forward, nearly-coherent scattering component and treating
the remaining component as isotropic scattering with an
angle-independent energy loss.
We call this approximated Compton opacity the iso-Compton opacity:
see J98 for the exact prescription.
The assumption of angle-independent energy loss ensures
that the scattered fields also consist of line spectra.


     Next we compute $\gamma$-ray energy $E_{\ApJstyle i,j}$ 
and fractional emissivity $f_{\ApJstyle i,j}$ for each order
of scattering $i$ and line $j$ assuming an infinite,
homogeneous, isotropic medium:  see J98 for the procedure.
We then use the  $E_{\ApJstyle i,j}$'s and
$f_{\ApJstyle i,j}$'s to construct total, absorption, and
scattering mean opacities for each order.
We adopt emissivity-weighted mean opacities since they offer
a good prospect of accurately reducing the $\gamma$-ray transfer
to grey radiative transfer for the purposes of energy deposition
(J98).
The mean opacity
prescription is

\midinsert                

{

\baselineskip=18pt
\tabskip=50pt minus 50pt  
\newdimen\digitwidth
\setbox0=\hbox{\rm0}
\digitwidth=\wd0
\catcode\lq?=\active
\def?{\kern\digitwidth}

\leftline{{\bf TABLE 1.}\ Mean opacities for solar and
mean model W7 compositions}
\vskip .5\baselineskip\hrule\smallskip\hrule\medskip
\halign to \hsize{\hfil#\hfil  &\hfil#\hfil &\hfil#\hfil &\hfil#\hfil
                               &\hfil#\hfil ??
                               &\hfil#\hfil &\hfil#\hfil &\hfil#\hfil &\hfil#\hfil \cr
\noalign{
\hskip 2.625cm \hbox{Solar composition} \hskip 2.925cm
          \hbox{Mean model W7 composition} }
\noalign{\hskip 2.625cm
       ($\mu_{\ApJstyle e}=\mu_{\ApJstyle e}^{\ApJstyle\odot}
        =1.179$) \hskip 4.475cm
           ($\mu_{\ApJstyle e}=2.095$) }
\noalign{\medskip\hrule\medskip}
Order &$\bar E_{\ApJstyle i}$ &$\kappa_{\ApJstyle i}$
      &$\xi_{\ApJstyle i}^{\ApJstyle\rm a}$
      &$\xi_{\ApJstyle i}^{\ApJstyle\rm s}$
      &$\bar E_{\ApJstyle i}$ &$\kappa_{\ApJstyle i}$
      &$\xi_{\ApJstyle i}^{\ApJstyle\rm a}$
      &$\xi_{\ApJstyle i}^{\ApJstyle\rm s}$ \cr
\noalign{\smallskip}
\noalign{\smallskip}
      &(MeV)
      &$\left({\rm cm^{\ApJstyle 2}\,g^{\ApJstyle -1}}\right)$
      &
      &
      &(MeV)
      &$\left({\rm cm^{\ApJstyle 2}\,g^{\ApJstyle -1}}\right)$
      &
      &  \cr
\noalign{\medskip\hrule\medskip}
\noalign{\vskip .5\baselineskip}
\noalign{\centerline{\rm$^{\ApJstyle 56}$Co}}
\noalign{\smallskip}
 0 &$ 1.24226$ &$0  .0547$ &$0  .7870$ &$0  .2130$ &$ 1.24226$ &$0  .0318$ &$0
  .7829$ &$0  .2171$\cr
 1 &$0 .23486$ &$0  .1307$ &$0  .3570$ &$0  .6430$ &$0 .23995$ &$0  .0827$ &$0
  .4393$ &$0  .5607$\cr
 2 &$0 .15099$ &$0  .1637$ &$0  .2524$ &$0  .7476$ &$0 .15294$ &$0  .1306$ &$0
  .4775$ &$0  .5225$\cr
 3 &$0 .11294$ &$0  .1873$ &$0  .1976$ &$0  .8024$ &$0 .11403$ &$0  .2006$ &$0
  .5810$ &$0  .4190$\cr
 4 &$0 .09074$ &$0  .2052$ &$0  .1639$ &$0  .8361$ &$0 .09147$ &$0  .3028$ &$0
  .6826$ &$0  .3174$\cr
 5 &$0 .07607$ &$0  .2194$ &$0  .1413$ &$0  .8587$ &$0 .07660$ &$0  .4459$ &$0
  .7630$ &$0  .2370$\cr
\noalign{\medskip}
\noalign{\centerline{\rm$^{\ApJstyle 56}$Ni}}
\noalign{\smallskip}
 0 &$0 .53479$ &$0  .0807$ &$0  .5911$ &$0  .4089$ &$0 .53479$ &$0  .0494$ &$0
  .6230$ &$0  .3770$\cr
 1 &$0 .18845$ &$0  .1443$ &$0  .3077$ &$0  .6923$ &$0 .19557$ &$0  .1043$ &$0
  .4762$ &$0  .5238$\cr
 2 &$0 .12888$ &$0  .1749$ &$0  .2245$ &$0  .7755$ &$0 .13679$ &$0  .1573$ &$0
  .5330$ &$0  .4670$\cr
 3 &$0 .09966$ &$0  .1964$ &$0  .1797$ &$0  .8203$ &$0 .10694$ &$0  .2298$ &$0
  .6205$ &$0  .3795$\cr
 4 &$0 .08180$ &$0  .2128$ &$0  .1515$ &$0  .8485$ &$0 .08797$ &$0  .3328$ &$0
  .7054$ &$0  .2946$\cr
 5 &$0 .06961$ &$0  .2259$ &$0  .1326$ &$0  .8674$ &$0 .07468$ &$0  .4764$ &$0
  .7756$ &$0  .2244$\cr
} 
\medskip\hrule
\vskip\baselineskip

\baselineskip=12pt
\narrower
     {\bf NOTE.}---The solar composition is the solar system composition
of Anders and~Grevesse 1989.
The mean model~W7 composition (Thielemann~\etal 1986)
is the final mean composition after all radioactive species have decayed.
$\mu_{\ApJstyle e}$ is the mean atomic mass per electron
(e.g., Clayton 1983, p.~84).

}

\endinsert

$$  \kappa_{\ApJstyle i}^{\ApJstyle R}=
{\sum_{\ApJstyle j} \kappa_{\ApJstyle i,j}^{\ApJstyle R}f_{\ApJstyle i,j}
                E_{\ApJstyle i,j}
\over
\sum_{\ApJstyle j} f_{\ApJstyle i,j}E_{\ApJstyle i,j}
                                                }  \,\, , \eqno(\eqn)$$


\noindent where $\kappa_{\ApJstyle i,j}^{\ApJstyle R}$ is evaluated
at energy $E_{\ApJstyle i,j}$.
The superscript $R$, here and elsewhere in this paper, is a variable
that replaces a symbol designating a quantity as related to
total (blank), absorption (``a''), or scattering (``s'')
opacity. 
The $\kappa_{\ApJstyle i,j}^{\ApJstyle R}$'s are the sums of the
iso-Compton, photoelectric, and pair production opacities
or opacity components.


     The mean $\gamma$-ray energy for an order is given by
$$  \bar E_{\ApJstyle i}=
{\sum_{\ApJstyle j} f_{\ApJstyle i,j}E_{\ApJstyle i,j}
\over
\sum_{\ApJstyle j} f_{\ApJstyle i,j}            }  \,\, . \eqno(\eqn)$$
The mean energy is not actually used in the LS~procedure, but it is
a useful diagnostic.
It happens that on first scattering for both solar and metal-rich
compositions that the $^{\ApJstyle 56}$Co and
$^{\ApJstyle 56}$Ni mean energies are reduced by more than 
$80\,$\% and $60\,$\%, respectively.
This result (which depends on the iso-Compton opacity approximation)
shows that the importance of the nonzero order fields is
relatively small.

     Values for $\kappa_{\ApJstyle i}$ and, additionally,
$\bar E_{\ApJstyle i}$ and
$\xi_{\ApJstyle i}^{\ApJstyle R}
\equiv\kappa_{\ApJstyle i}^{\ApJstyle R}
 /\kappa_{\ApJstyle i}$ 
($i\in[0,5]$)
for $^{\ApJstyle 56}$Co and $^{\ApJstyle 56}$Ni in solar
and mean model~W7 compositions are given in Table~1.


\section{The LS procedure}
What we want from a grey radiative transfer procedure is
the $\gamma$-ray energy deposition as function of position.
Since supernova density varies widely with 
position and epoch, it is convenient to measure
energy deposition by $\epsilon_{\ApJstyle\rm d}$,
the energy deposited per unit time per unit mass.
The LS~procedure expression for the energy deposition at a point is
$$ \epsilon_{\ApJstyle\rm d}
 =4\pi\kappa_{\ApJstyle\rm eff}^{\ApJstyle\rm a} J_{\ApJstyle 0}
                                               \,\, , \eqno(\eqn)$$
where $\kappa_{\ApJstyle\rm eff}^{\ApJstyle\rm a}$ is the
effective absorption opacity and $J_{\ApJstyle 0}$ is the
0th order frequency-integrated mean intensity (or radiation field).
The $J_{\ApJstyle 0}$ field is calculated from an numerical
integral solution of the radiation transfer equation 
using the 0th order frequency-integrated source function
(determined by the radioactive species) and the 0th order
mean opacity $\kappa_{\ApJstyle 0}$.

     The expression for
$\kappa_{\ApJstyle\rm eff}^{\ApJstyle\rm a}$ is
$$ \kappa_{\ApJstyle\rm eff}^{\ApJstyle\rm a}=
    \kappa_{\ApJstyle 0}L \,\, ,
\qquad{\rm where}\qquad
 L=
\left(\sum_{\ApJstyle i=0}^{\ApJstyle k-1}
     \xi_{\ApJstyle i}^{\ApJstyle\rm a}
      \prod_{\ApJstyle j=0}^{\ApJstyle i-1}
       \xi_{\ApJstyle j}^{\ApJstyle\rm s}
        \zeta_{\ApJstyle j+1}\right)
+
\left(\prod_{\ApJstyle j=0}^{\ApJstyle k-1}
       \xi_{\ApJstyle j}^{\ApJstyle\rm s}
        \zeta_{\ApJstyle j+1}\right)
{ \xi_{\ApJstyle k}^{\ApJstyle\rm a} \over
   1-\xi_{\ApJstyle k}^{\ApJstyle\rm s}
        \zeta_{\ApJstyle k} }      \,\,        \eqno(\eqn)$$
is a series accounting for absorption from all orders.
The $\zeta$ factors in equation~(\showeqn -0) are
defined by
$$\zeta_{\ApJstyle i}=\oint{d\Omega\over 4\pi}\,
       \exp\left[1-\exp\left(-\tau_{\ApJstyle i}\right)\right]
       \approx
{1\over2}\left\{
\left[1-\exp\left(-\tau_{\ApJstyle i,{\rm out}}\right)\right]
+\left[1-\exp\left(-\tau_{\ApJstyle i,{\rm in}}\right)\right]\right\}
                                             \,\, ,  \eqno(\eqn)$$
\smallskip
\noindent where $\Omega$ is solid angle and
$\tau_{\ApJstyle i}$ is the $i$th order mean (total) opacity optical
depth from the deposition point to the surface of the supernova.
The second expression in equation~(\showeqn -0) is a two-stream
approximation to the first for spherically symmetric cases:
$\tau_{\ApJstyle i,{\rm out}}$ and $\tau_{\ApJstyle i,{\rm in}}$
are the outward and inward radial $\tau_{\ApJstyle i}$ values.

     We derived equation~(\showeqn -1) making the LS approximation
for the nonzero order fields by assuming the nonzero order
fields could be
approximated by their values at the deposition point itself.
We also assumed that all the mean opacities from the $k$th
order on could be approximated by the $k$th mean opacity.
Using this assumption we summed the contributions from the $k$th term
on analytically in the $L$ series.
Since most of the deposition comes from the lowest order fields
(and thus the lowest order terms in the $L$ series), the crude
approximation for terms $>k$ is adequate.
For the cases we have studied, the difference in global energy
deposition between choosing $k$ to
be 2 and 5 turned out to be $\lapprox 0.6\,$\%.
We assume $k=5$ to be our fiducial $k$ value.

     In summary, the whole LS~procedure consists of doing the
numerical $\gamma$-ray transfer for the 0th order field and then
evaluating the effective absorption opacities
and the energy deposition.

\section{Conclusions}
In this paper, we have outlined the LS~procedure for
grey $\gamma$-ray transfer and energy deposition
in supernovae.
A detailed derivation is given by J98.
The LS~procedure has the advantage over previous like
procedures in that it has no free parameter.

     J98 concluded that the LS~procedure is
modestly more reliable than the SSH procedure sans 
an optimized mean opacity and estimated that the
maximum uncertainty in the LS~procedure for global
energy deposition may be of the order $6\,$\%.
An extensive comparison of LS~procedure results to
those of a more accurate procedure is 
needed to definitively assess the accuracy of the 
LS~procedure. 

     A computer code for the LS~procedure can be obtained by
request from the author.

\ack
   This work was supported by the U.S.~Department of Energy's
   Office of Fusion Energy Sciences
   under Contract No.~DE-AC05-96OR22464 with
   Lockheed Martin Energy Research Corp.~and by the
   ORNL Research Associates Program administered jointly
   by ORNL and the Oak Ridge Institute for Science and Education.
   I thank Peter Sutherland for providing me with his
   grey radiative transfer code for $\gamma$-ray energy deposition
   and the symposium organizers for their mighty deeds.

\vskip-14pt
\references
Ambwani K and Sutherland P G 1988 {\it Astrophys.~J.} {\bf 325} 820\par
Anders E and~Grevesse N 1989 {\it Geochim.~Cosmochim.~Acta}
              {\bf 53} 197\par  
Browne E and Firestone R B 1986 {\it Table of Radioactive
          Isotopes} (New York:  John Wiley \& Sons, Inc.)\par
Chan K W and~Lingenfelter R E 1993 {\it Astrophys.~J.} {\bf 405} 614\par
Clayton D D 1983 {\it Principles of Stellar Evolution and
              Nucleosynthesis} (Chicago:  Univ.~of Chicago Press)\par
Colgate S A, Petschek A G, and~Kriese J T 1980, {\it Astrophys.~J.}
            {\bf 237} L81\par
Davisson C M 1965 {\it Alpha-, Beta-, and Gamma-Ray Spectroscopy}
             ed K~Siegbahn (Amsterdam:  North-Holland) p 37\par
Huo J 1992 {\it Nuclear Data Sheets} {\bf 67} 523\par
Jeffery D J 1998 preprint (J98)\par
Milne P A, The L-S, and Leising M D 1997,
             {\it Proc.~the Fourth Compton Symposium, Williamsburg,
             Virginia} ed C~D~Dermer~\etal
             (New York:  American Institute of Physics Press) p~1022\par
Ruiz-Lapuente P 1997 {\it Proc.~NATO ASI on
             Thermonuclear Supernovae, Begur, Girona (Spain)}
             ed P Ruiz-Lapuente~\etal
             (Dordrecht:  Kluwer)\par
Sutherland P G and Wheeler J C 1984 {\it Astrophys. J.} {\bf 280} 282\par
Swartz D A, Sutherland P G, and~Harkness R P 1995
            {\it Astrophys.~J.} {\bf 446} 766 (SSH)\par
Thielemann F-K, Nomoto K, and~Yokoi K 1986, {\it Astr.~Astrophys.}
             {\bf 158} 17\par
\endrefs

\bye